\documentstyle[12pt]{article}

\newcommand{\be}{\begin{equation}}
\newcommand{\ee}{\end{equation}}
\newcommand{\beqn}{\begin{eqnarray}}
\newcommand{\eeqn}{\end{eqnarray}}

\begin{document}
\begin{center}
\begin{large}
{\bf Flavor Singlet Contribution to the Structure Function $g_1$ \\
at Small-x} \\
\vspace{1cm}
J. Bartels\footnote
        {Supported by Bundesministerium f\"ur Forschung und
        Technologie, Bonn, Germany under Contract 05\,6HH93P(5) and
        EEC Program "Human Capital and Mobility" through Network
        "Physics at High Energy Colliders" under Contract
        CHRX-CT93-0357 (DG12 COMA)} \\
\end{large}
{\it II. Institut f\" ur Theoretische Physik, Universit\" at Hamburg}
\\
\vspace{1cm}
\begin{large}
B.I.Ermolaev\footnote{The research described in this paper has been
made possible in part by the Grant R2G 300 from the International
Science Foundation and the Russian Government. This work has also been
supported by the Volkswagen-Stiftung.}
\\
\end{large}
{\it A.F.Ioffe Physical-Technical Institute, St.Petersburg,
194021, Russia}
\\
\vspace{1cm}
\begin{large}
 M. G. Ryskin\footnote{Work supported by the Grant INTAS-93-0079
and by the Volkswagen-Stiftung}
\end{large}
\\
{\it Petersburg Nuclear Physics Institute, Gatchina, St.Petersburg,
188350, Russia}
\\
\end{center}
\vspace{1cm}
{\bf Abstract:}
\noindent
The singlet contribution to the $g_1(x,Q^2)$ structure function is
calculated in the double-logarithmic approximation of perturbative
QCD in the region $x \ll 1$. Double logarithmic contributions
of the type $(\alpha_s \ln ^2 (1/x))^k$ which are not included in
the GLAP evolution equations are shown to give a power-like rise
at small-x which is much stronger 
than the extrapolation of the GLAP expressions. 
The dominant contribution is due to the gluons which, in contrast to the
unpolarized case, mix with the fermions also in the region $x \ll 1$.
The two main reasons why the small-x
behavior of the double logarithmic approximation is so much stronger than
the usual GLAP evolution are: the larger kinematical region of integration
(in particular, no ordering in transverse momentum) and the contributions
from non-ladder diagrams.
\newpage
\section{Introduction}
\setcounter{equation}{0}
The investigation of the structure functions $g_1$ and $g_2$ provides the
basis for the theoretical description of polarization effects in deep
inelastic lepton nucleon scattering. Particular interest has been given to
the behavior of $g_1$ at small x: experimental data ~\cite{data} 
on both $g_1^P$ and
$g_1^N$ are limited to $x>10^{-2}$, and numerical values of the moments
$\Gamma^{P,N}=\int_0^1 dx g_1^{P,N}$ therefore depend upon the extrapolation
in the region $x<10^{-2}$.\\ \\
In QCD the $Q^2$ evolution at not too small x 
of $g_1$ is, like in the unpolarized case,
described by the GLAP-evolution equations ~\cite{GL,AP}. The prediction 
for the small-x behavior is that both the quark and the gluon 
polarized structure functions go as
\beqn
\Delta q(x,Q^2), \Delta g(x,Q^2) \sim \exp \sqrt{ 
const \cdot \alpha_s \ln(Q^2/\mu^2) \ln (1/x)}.
\eeqn
In particular, in the flavor singlet part the gluons and the quarks mix.
In a recent paper ~\cite{BER} it has been shown, for the flavor nonsinglet
contribution to $g_1$, that this simple double log extrapolation (1)
of the GLAP evolution equations, in fact, strongly underestimates the
rise at small x. The reason is that, at $x \ll 1$, new double logarithmic 
contributions appear which are beyond the control of the standard evolution
framework. To be more precise, in the region of not too small x where
the standard analysis applies the leading 
behavior in the n-th order of $\alpha_s$ is of the type
\beqn
(\alpha_s \ln(Q^2/\mu^2))^n [a_n ( \ln (1/x) )^n + a_{n-1}( \ln(1/x)^{n-1}
                       + ...]
\eeqn
(where $\mu$ denotes the renormalization scale, and for simplicity we restrict
ourselves to a fixed $\alpha_s$),
whereas at very small x the dominant contributions are of the form:
\beqn
(\alpha_s \ln(1/x) \ln(1/x) )^n.
\eeqn
These double logarithmic contributions are not included in the standard
evolution scheme ~\cite{AP}. In ~\cite{BER} the sum of the double logarithms 
\beqn
(\alpha_s)^n [b_n (\ln (1/x))^{2n} + b_{n-1} (\ln(1/x))^{2n-1} \ln (Q^2/\mu^2)
               +...+b_0 (\ln(1/x))^n (\ln(Q^2/\mu^2))^n
\eeqn
has been shown to give rise to a power-like increase 
of the flavor nonsinglet structure function at small-x which
is stronger than the GLAP predicition (1).\\ \\
In this paper we continue our investigation of the small-x behavior of
$g_1$ in the double logarithmic approximation, by calculating the flavour
singlet contribution. Like in the flavor nonsinglet case, we derive and solve
evolution equations which describe the dependence upon the infrared cutoff
of the transverse momentum integrations. Our main results is the power-like
growth of $g_1$ at small-x (eqs.(4.21) - (4.23)). A discussion of the 
phenomenological
implications of our result will be presented in a forthcoming paper.\\ \\

\section{Singlet contribution to $g_1$ in lowest orders  $\alpha_s$}
\setcounter{equation}{0}

Let us consider the scattering process where the off-shell photon with 
momentum $q$ ($-q^2=Q^2>0$) strikes the polarized quark with momentum $p$. 
We assume that the total energy is much greater than the mass of the quark,$m$.
\be
s=2pq>>m^2.
\label{f5}
\ee 
We follow the notation of ~\cite{BER,IKL} and introduce the scattering 
amplitudes $T_i$ in the following way:
\beqn
T_{\mu \nu}& = & \imath \int d^4x e^{\imath
qx}       <N|T \left(J_{\mu} (x) J_{\nu}(0) \right)| N>
   \:\:\:\:\:\:\:\:\:\:\:\:\:\:\:\:\:\:\:\:\:\:\:\:\:\:\:\:\:\:\:\:\:
       \nonumber \\
           & = &
    (-g_{\mu \nu} + \frac{q_{\mu} q_{\nu} } {q^2})T_1
  + (p_{\mu} - q_{\mu} \frac{pq}{q^2} )
    (p_{\nu} - q_{\nu} \frac{pq}{q^2} ) T_2
         \nonumber \\
&   &   + \imath \epsilon_{\mu \nu \alpha \beta} q^{\alpha} s^{\beta}
           \frac{m}{pq} T_3
  + \imath \epsilon_{\mu \nu \alpha \beta} q^{\alpha}
             (s^{\beta}(pq) - p^{\beta} (sq) ) \frac{m}{pq} T_4.
\eeqn
The structure function $g_1$ is defined as
\beqn
g_1 =- \frac{1}{2 \pi} Im T_3.
\eeqn
\\ \\
In the Born approximation the matrix element for this process is given by
the Feynman graphs in Fig 1. Let us denote by $M_L (M_R)$ the matrix element
for the case when the initial quark has left (the right) helicity. Then
we obtain
\beqn
T_3^{Born}=\frac{M^{Born}_L-M^{Born}_R}{2} 
          =e^2_q
        (\frac {s}{s-Q^2+\imath\epsilon}-\frac {-s}{-s-Q^2+\imath \epsilon})
\label{f6}
\eeqn
so that
\be
g^{Born}_1=-\frac{1}{2\pi}Im M^{Born}.
\label{f7}
\ee
$e_q$ here stands for the electric charge of the initial quark. Expession 
(\ref{f6}) shows that $T_3^{Born}$ is antisymmetrical with respect 
to replacing $s$ by $-s$, i.e. in terms of the Regge theory $T_3^{Born}$ 
has negative/odd signature.
QCD radiative corrections turn $T_3^{Born}$ into $T_3$. As usual, it consists
of the flavor nonsinglet part $T_3^{NS}$ and the flavour singlet part
$T_3^{S}$. The small-x behavior of the nonsinglet contribution has 
been discussed in ~\cite{BER}, and in this paper we shall consider the singlet
part. All Feynman graphs contributing to the double-logarithmic approximation
(DLA) of the nonsinglet part $T_3^{NS}$ can be obtained from the 
graphs in Fig.1 by adding the gluon lines (further rungs and
nonladder bremsstrahlungs gluons). An example is shown in Fig.2. 
In the present case of the flavor singlet we will have to include more
general contributions, in particular gluon ladders. An example is shown in
Fig.3. As a result, we first have to consider the four different types of 
kernels: $\Delta P_{qq}$, $\Delta P_{qg}$, $\Delta P_{gq}$, and 
$\Delta P_{gg}$. They will be derived in the remainder of this 
section and agree with the limit $x \to 0$ of the leading order 
splitting functions of ~\cite{AP}. In addition to these 
expressions for the kernels, we find the region of integration which gives the
double logarithmic contributions. Next we will have to consider the
bremsstrahlungs gluons. As the result of this section, we obtain all the 
ingreduents for the infrared evolution equations that will be discussed in the
following section. Throughout this paper we will use the Feynman gauge 
and the Sudakov representation for the quark and gluon momenta:
\be
k_1=\alpha_i q' +\beta_i p +k_{Ti},
\label{f11}
\ee
\be
q'=q+xp,
\label{f12}
\ee
and
\be
dk_i=\frac{s}{4}d \alpha_i d \beta_i d k^2_{Ti} d \varphi_i.
\label{f13}
\ee
\\ \\
For the beginning let us briefly review ~\cite{BER} the structure of the double
log contribution in the non-singlet case. The imaginary part of the
amplitude $T_{3}$ comes from the $s$-channel graph Fig.1, and the spin
structure of ($M_{L}$--$M_{R}$), corresponding to the quark line has
the form 
\beqn 
tr \left[
\gamma_{5}\hat{p}\gamma_{\mu}(\hat{q}+\hat{p})\gamma_{\nu} \right] =
-4i\varepsilon_{\mu\nu\alpha\beta} q^{\alpha}p^{\beta}, \,\,\,
        \left(
        \gamma_{5}=-i\varepsilon_{\alpha\beta\gamma\delta}
        \gamma^{\alpha}\gamma^{\beta}\gamma^{\gamma}\gamma^{\delta}
        \right).
\eeqn 
which
coincides with the tensor $i\varepsilon_{\mu\nu\alpha\beta} q^{\alpha}
s^{\beta}\cdot m$ in eq.(2.2) for the longitudinal spin vector
$s^{\beta}=p^{\beta}/m$ of the initial quark.  To obtain the leading
logarithm in the first loop Fig.2a, one has to cancel one of the two
transverse momenta factors $k_{t}^2$ from the $t$-channel quark
propagators. In the trace 
\beqn 
tr \left[ \gamma_{5}\hat{p}\gamma_{\alpha'}\hat{k}\gamma_{\mu}(\hat{q} +
\hat{k})\gamma_{\nu} \hat{k}\gamma_{\alpha'} \right] \simeq 2tr \left[
\gamma_{5}\hat{p}\hat{k}\gamma_{\mu}\hat{q}'\gamma_{\nu}\hat{k}
\right] 
\eeqn 
we have to keep the largest momentum $q'$ in the
$s$-channel quark propagator $(\hat{q}+\hat{k})$, and retain only the 
transverse
components of the photon polarization vectors ($\mu$ and $\nu$) (see
eq.~(2.9)).  Therefore one can omit the $\alpha q'_{\mu}$ component of
$k_{\mu}$ 
\beqn 
 tr \left[ \dots \alpha\hat{q'}\hat{q'}\dots
\right]=\alpha q'^{2}tr \left[ \dots \right] =0 \;\;\;\; as\;\;\;\;
q'^{2}=0 
\eeqn 
and write 
\beqn 
2tr \left[
\gamma_{5}\hat{p}\hat{k}\gamma_{\mu}\hat{q}'\gamma_{\nu}\hat{k}
\right] = 2k^{2}_{t} tr \left[
\gamma_{5}\hat{p}\gamma_{\mu}\hat{q}\gamma_{\nu} \right]. 
\eeqn
Together with the quark propagators $(1/k^{2}_{t})^{2}$ and the colour
factor $C_{F}=(N^{2}_{c}-1)/2N_{c}$ it gives the logarithmic loop
integral: 
\beqn 
{2C_{F}\over 4\pi} \alpha_s {dk^{2}_{t}\over k^{4}}\cdot
k^{2}_{t}
\eeqn 
The same reasoning is valid for the second loop fig.2b \cite{GGFL}.  As a result, 
we obtain for the gluon rung inside the fermion ladder $\Delta P_{qq}$:
\beqn
\Delta P_{qq}=C_F.
\eeqn
This result is also valid for the flavor singlet.\\ \\
Turning now to the singlet case, in the second loop approximation a new 
type of diagrams appears, contributions with $t$-cannel gluons.
The first diagram of this type, Fig.3, contains two fermion traces. The lower
one, originating from the initial quark spinors, has the form eq.(2.9) with
the gluon momentum $(-k^{\alpha})$ instead of $q^{\alpha}$. It
is antisymmetric in the polarizations of the two gluons with
momentum $k_1$. Thus we cannnot assign the same (longitudinal) polarizations to
both gluons: otherwise we would have $M_{R}=M_{L}$ and 
$T_{3}={1\over   2}(M_{L}-M_{R})=0$. The best one can do in order to retain
the largest power of $1/x$ (in the small $x$ region) is to assign the 
longitudinal polarization  $e_{\mu}\propto q'_{\mu'}$ to one of the two
gluon and the transverse polarization $(e_{\nu}'=e_{\nu't}$) to the other
gluon. In the proton (quark) rest frame with $k_{1t}=k_{x}$ this means that
$e_{\mu '}=e_{z}$, $e_{\nu '}=e_{y}$, and the lower trace
\beqn
tr_{l}=-tr
\left(
\gamma_{5}\hat{p}\hat{e}_{\mu'}\hat{k}_{1}\hat{e}_{\nu'}
\right)
=
4i\varepsilon_{\mu'\nu'\alpha\beta} k^{\alpha}_{1t}p^{\beta}
=
-4ie_{z\mu'}|p\parallel k_{1t}|e_{y\nu'}.
\eeqn
To obtain the logarithmic integration over $k_{1}$ we have to find
another factor $k_{1t}$ from the upper trace, i.e. to take the
component $\hat{k}_{1t}$ in the quark propagator
$(\hat{k}_{2}-\hat{k}_{1})$~--- the only one, which contains the
momentum $k_{1t}$. 
The product of three $\gamma$-matrices $\hat{e}_{\mu'}$,
$\hat{k}_{1t}\hat{e}_{\nu'}$, in the upper trace may be written as
\beqn
       \hat{e}_{\mu'z}\hat{k}_{1tx}\hat{e}_{\nu'y}
=
        -i\gamma_{5}    \gamma_{0}|k_{1t}|\quad .
\eeqn
Therefore the trace corresponding to the fermion loop in Fig.3
 takes the same form (2.10) as for the Fig.2b
\beqn
        \sim tr
        \left[
        \gamma_{5}\hat{p}\hat{k}_{2}\gamma_{\mu}\hat{q}\gamma_{\nu}\hat{k}_{2}
        \right],
\eeqn
and in this way the polarization of the target quark is
transfered to the fermion loop. So the upper quark loop can be
calculated in analogy with the one-loop graph in Fig.2a. As
usual, the DL-contribution comes from the kinematical region of
\beqn
        1\gg \beta_{1}\gg \beta_{2}\gg\dots
\eeqn
and
\beqn
       \dots \ll \alpha_{1}\ll \alpha_{2}\ll 1.
\eeqn
where all the $s$-channel (horizontal lines) particles are on mass
shell.\footnote{The ordering $1>\beta_{i}>\beta_{i+1}>0$,
$0<\alpha_{i}<\alpha_{i+1}<1$ arises from the energy-momentum
conservation, and within the DLA one can change it by the strong
ordering eq.(2.13); this is the largest region which may give the
DL-contribution.} In this case $\alpha_{i}\beta_{i}s\ll k^{2}_{it}$,
so $k^{2}_{i}\approx k^{2}_{it}$.\\ \\
Finally, the double logarithmic contribution of the diagram Fig.3 to the $g_{1}$
structure function is:
\begin{equation}
M_{3a}
=
 - \left(         \sum\limits^{n_{F}}_{q}e^{2}_{q}  \right)    2C_{F}
        \left(   {\alpha_{s}\over 2\pi}   \right)^{2}
        \int\limits^{1}_{x}
        {d\beta_{1}\over\beta_{1}}
        \int{d^{2}k^{2}_{t1}\over k^{2}_{t1}}
        {dk^{2}_{t2}\over k^{2}_{t2}}.
\end{equation}
From this expression we extract the results for the transitions 
$\Delta P_{gq}$ and $\Delta P_{qg}$:
\beqn
\Delta P_{gq} = 2 C_F, \\
\Delta P_{qg} = -2T_f.
\eeqn
\\ \\
It is interesting to note that the element corresponding to the gluon to
quark transition (i.e. $\Delta P_{qg}$ splitting kernel) $-2T_f$ 
is negative. From the physical point of view it means that the slow
(low x) produced quark has the helicity opposite to the one of 
the parent gluon; the initial gluon helicity goes to 
the fast antiquark with larger x. \\ \\
At this point it may also be worthwhile to stress the change in the energy 
dependence of $g_1$ compared to the unpolarized case, where the longitudinal 
polarization of both t-channel gluons leads to the behavior
$f_1(x)\propto 1/x$, i.e. to a singularity at angular momentum $j=1$.
In the present case we cannot have two longitudinally polarized t-channel
gluons and therefore loose one power in $1/x$: 
\beqn
 g_{1}(x)\sim x^{0}\cdot F(ln(1/x))\;\;\; at\;\;\; x\to 0.
\eeqn
Consequently, the rightmost singularity of the function $g_{1}$ in the 
complex angular momentum $j$-plane lies at $j=0$.
The $t$-channel quark exchange gives the same power of $1/x$ at
small $x$, and therefore quark and gluon structure functions mix at small
x. This is in contrast to the unpolarized case, where at small x the gluons
dominate by one power in $1/x$.\\ \\
To complete our discussion of the rungs, we have to adress the 
diagrams with gluon rungs in a gluon ladder. Such a contribution appears 
first in the $3^{d}$ loop, and it is illustrated Fig.4. To
simplify the kernel we will write the triple-gluon vertex
as a sum of three pieces (Fig.5a):
\beqn
        \Gamma_{\mu\nu\rho}
=
        g_{\mu\nu}(k_{1}+k_{2})_{\rho}
+
        g_{\mu\rho}     (k_{2}-2k_{1})_{\nu}
+
        g_{\nu\rho}(k_{1}-2k_{2})_{\mu},
\eeqn
and note the following:\\
1) due to the antisymmentry in the indices $\mu'\nu'$ the graphs
5b,c do not contribute to $g_{1}$ (Fig.4);\\
2) due to gauge invariance we have
$M_{\mu'\nu'}k_{1\mu'}=M_{\mu'\nu'}k_{1\nu'}=0$ and
$N_{\mu''\nu''}k_{2\mu''}=N_{\mu''\nu''}k_{2\nu''}=0$;\\
3) graph Fig.5d does not give the leading logarithm, as the product $(k_{1}
+k_{2})^{2}=2(k^{2}_{1}+k^{2}_{2})$ 
(note that the $s$-channel gluon
$k_{1}-k_{2}$ is on shell and $(k_{1}-k_{2})^{2}=0$; the term $\sim
k^{2}_{1}$ (or $k^{2}_{2}$) cancels one of the gluon propagators and
kills the logarithm $\frac{dk^{2}_{t}}{k^{2}_{t}}$ in the integral over
$k_{1t}$ (or $k_{2t}$) ). 
Thus the product of two triple gluon vertices in Fig.4 takes the form:
\beqn
H_{\mu'\nu',\mu"\nu"}
=
4\Bigl(
        \delta_{\mu'\mu"} k_{2\nu'}k_{1\nu"}
+
        \delta_{\nu'\nu"}k_{2\mu'} k_{1\mu"} -
\delta_{\mu'\nu"}k_{2\nu'}k_{1\mu"}
-
        \delta_{\nu'\mu"}k_{1\nu"}k_{2\mu'},
\Bigr)
\eeqn
which corresponds to the diagrams Fig.5c.\\ \\
Together with the colour coefficient $N_{c}$ it means that the
insertion of an extra $s$-channel gluon into the gluon loop of Fig.3
leads to the double log integration
\begin{equation}
        4N_{c}{\alpha_{s}\over 2\pi}\int
        {d\beta_{2}\over \beta_{2}}
        \int{dk^{2}_{t}\over k^{2}_{t}}.
\end{equation}
Indeed, in order to keep the largest power of $1/x$ and to save the logarithm
in the integrals over $k_{t}$ we have to choose in (2.25) the
transverse components of $k_{1}$ and $k_{2}$ and the longitudinal
indices in $\delta_{\mu'\mu"}$, $\delta_{\nu'\nu"}$,
$\delta_{\mu'\nu"}$ or $\delta_{\nu'\mu"}$. Such a
configuration conserves the main structure of the gluon loop
fig.~3, inserting instead of the spin part of the transverse
gluon propagator $-e_{\nu'y}e_{\nu"y}$ (in Fig.~3) the
expression $(e_{\nu"}k_{1t})(e_{\nu'}k_{2t})=-|k_{1t}\parallel
 k_{2t}|\sin^{2}\varphi$
(here $\varphi$ is the angle between the transverse momenta $k_{1}$
and $k_{2}$ and we take into account the fact that $e_{\nu'}\perp
k_{1t}$ and $e_{\nu''}\perp k_{2t}$).
An additonal power of $|k_{1t}|$ and $|k_{2t}|$ comes from the traces
of the lower and upper quark loops (like in the case of eq.(2.15)) and
(2.16)). So after the integration over the azimuthal angle $(\langle
\sin^{2}\varphi\rangle=1/2)$ the vertex eq.(2.25) gives us the result (2.26)
From this we extract the gluon rung $\Delta P_{gg}$:
\beqn
\Delta_{gg}=4N_c = 4C_A
\eeqn
Finally, let us collect our results for the four different rungs.
We define a matrix $M_0$ as illustrated in Fig.6, which contains the
splitting functions $\Delta P_{ij}$:
\beqn
M_0=
       \left( \begin{array}{cc}
               4C_A & -2T_f \\
               2C_F  & C_F
               \end{array}  \right)
\eeqn
(here $C_A=N$, $C_F=\frac{N^2-1}{2N}$, and $T_f= \frac{n_f}{2}$ are the 
usual $SU(N)$ color factors; note that we have chosen to put the gluons 
into the first column and row).
This matrix will be used in the following section where we shall derive the 
infrared evolution equations. For later purposes it will be convenient to
consider also the color octet t-channel. In this case the color matrix
analogous to (2.28) reads:
\beqn
M_8 = \left( \begin{array}{cc}
          2C_A & -T_f \\
           C_A & -1/2N
           \end{array}  \right).
\eeqn
\\ \\
From our previous discussion we have also obtained the general pattern 
of the region of phase space which gives the double logarithmic 
contributions.The limits of integrations follow from the ordering condition 
given in eqs.(2.18),(2.19). In terms of $\beta_{i}$ and $k_{ti}$ it means that
\cite{EMR,GGFL} 
\beqn
k^{2}_{t,i+1}\gg k^{2}_{t,i}{\beta_{i+1}\over\beta_{i}}\quad\quad
        (here\;\;\; k^{2}_{ti}>0).
\eeqn
Indeed, let us first take $\beta_{i+1}\ll \beta_{i}$ and $k_{t,i+1}>k_{t,i}$.
In this case $\alpha_{i+1}\approx {k^{2}_{t,i+1}\over
s\beta_{i}}\gg \alpha_{i}\approx {k^{2}_{t,i}\over s\beta_{i-1}}$ (we have
used the fact that $s$-channel partons with the momenta
$k_{i}-k_{i+1}$ are on mass shell; $(k_{i}-k_{i+1})^{2}\approx
-k^{2}_{t,i+1}-\alpha_{i+1}\beta_{i}s=0$). On the other hand, if
$k_{t,i+1}<k_{t,i}$ one has $\alpha_{i+1}\approx
{k^{2}_{ti}\over s\beta_{i}}$. In order to save the leading logarithm, we 
have to satisfy the condition: $k^{2}_{i+1}\approx -k^{2}_{t,i+1}$, ~i.~e.
$\alpha_{i+1}\beta_{i+1}s\ll k^{2}_{t,i+1}$. In other words, our
condition looks as
\begin{equation}
\tilde{\alpha}_{i+1}=\frac{k^{2}_{t,i+1}}{\beta_{i+1}s}\gg
\frac{k^{2}_{ti}}{\beta_{i}s}\equiv \tilde{\alpha}_{i}
\end{equation}
and as in \cite{GGFL} we can simply use the ordering eq.(2.19) with the
$\tilde{\alpha}_{i}$ (eq.(2.31)) instead of the $\alpha_{i}$.\\ \\
In the final part of this section we have to consider nonladder 
diagrams as illustrated in Fig.7. We will call a 
non-ladder gluon "soft" if its transverse momentum is
smaller than the momenta of all the partons comprised by the non-ladder
gluon. According to Gribov's $k_T$-factorization theorem \cite{Gri}
\footnote{For QCD the Gribov's theorem was considered
 in more detail in \cite{KL,Gri,EL} and \cite{CE} } 
the whole amplitude of the 
'soft' gluon emission can be written as the bremsstrahlung from one of 
the ``external`` lines of the block comprized by this gluon. 
Therefore the double logarithmic 
contribution coming from such a 'soft' gluons can be summed up with
the help of the infrared evolution equation\cite{KL}, in the same way as 
it was done for the non-singlet structure function $g_1$ in\cite {BER}.
When summing over all possibilities of attaching the soft gluon to the 
external legs, we get a total color factor which depends on both the total
t-channel color quantum number and the type of incoming partons. We will 
need the color singlet channel. For incoming gluons and fermions the 
color factors are $C_A$ and $C_F$, resp. In matrix notation we define
\beqn
G_0=\left( \begin{array}{cc}  C_A & 0 \\ 0 & C_F \end{array} \right).
\eeqn
\\ \\
Finally we note that 
non-ladder gluons with $k'_t$ larger than the transverse momenta in 
the part of the ladder, which is comprised by them  (they will be called 
'hard') do not give double logarithms. For a nonladder gluon that runs across
the ladder from one side to the other (e.g. from the lower left to the upper
right leg), the large momentum has to flow through some internal 
small $k_{ti}$ ladder propagators, and the large
momentum $k'_{t}$ changes the normal $1/k^{2}_{ti}$ factor to
$1/k'^2_{t}$, in this way killing the leading logarithm
$dk^{2}_{ti}/k^{2}_{ti}$ (see Fig.7). These hard nonladder
gluons therefore do not contribute to the double logarithmic approximation.
Next we consider  vertex correction (Fig. 7b). 
Then we can say that in the ultraviolet (large $k'_t$) 
 region there are no any double logs in the Feynmann gauge for the vertex
 function. So, one can
anticipate that there are no $DL$-correction coming from
large-$k_t$-region in the Feynmann gauge for the vertex function.
However, we have to be more carefull here. For the unpolarized case
there exists an example (BFKL) where a special cutting of the
vertex-type diagram does give rise to a double log contribution. Indeed, let
us consider the non-ladder on-shell gluon $k'$ added to the
amplitude: the loop integration yields ${d^{2}k'_{t}d\beta'\over
16\pi^{3}\beta'}$, two propagators~--- $1/q^{2}\approx
{1\over\alpha_{q}\beta's}$, $1/k^{2}_{2}\sim 1/k^{2}_{2t}$ and the
spin part of the propagators, which for the nonsense, longitudinally
polarization ($g^{n}_{\nu\nu'}\approx Q'_{\nu}p_{\nu '}/(Q'p)$) in the DL
kinematical region $\beta_2\ll\beta'\simeq\beta_{1}$
($\alpha_{q}\gg\alpha'$) gives the factor $\simeq
-(qp)(2k_{1}Q')/(pQ)\simeq \alpha_{q}\beta's$. This factor cancels
the propagator $1/q^{2}$, while the next propagator $1/k^{2}_{2t}\sim
{1\over k'^{2}_{t}}$ provides the logarithmic integration
$dk'^{2}_{t}/k'^{2}_{t}$ in the region $k'_{t}>k_{1t}$. As it is
known \cite{BFKL}, the sum of such contributions is equal to the ladder
contribution (Fig.~4) but has an opposite sign. So it cancels the
double logs coming from $k'_{t}>k_{ti}$ and restores the conventional
DGLAP ordering $k_{t,i+1}>k_{t,i}$ for the gluon loops.
Fortunately,  this does not happen for the spin dependent structure function
$g_{1}$. In this case the vertex of the "nonsense" gluon $k'$
emission changes the sign, if unstead of the
longitudinal polarizations of $t$-channel gluons $k_{1}$ and $k_{2}$
(in the left side of Figs.7c,d we consider the transverse
polarization $e^{t}_{1}\perp k_{1t}$.\footnote{To save
logarithm in $d\beta'/\beta'$ integration the gluon $k'$ should be
the longitudinal, nonsence one; on the other hand the polarization
vectors of the $t$-channel gluons $k_{1}$ (or $k_{2}$) and
$\tilde{k}_{1}$ should be~--- one transverse and one longitudinal
(for the small-x limit of $g_{1}$) as it was discussed just before
eq.~(11). If the vector $e_{1}\parallel p$ than the leading
contribution comes from the graphs Fig.7e,f and
$\Gamma_{\mu\nu\rho}\cdot p_{\mu}Q'_{\nu}Q'_{\rho}/(pQ')\simeq
-{k_{1}Q'\over pQ'}=-\alpha_{1}$, while for the transverse vector
$e_{1}=e_{1t}=e_{2t}$ only the diagram Fig.7e do work and
$\Gamma_{\mu\nu\rho}e_{1t\mu}e_{2t\nu}Q'_{\rho}/pQ'\approx
+\alpha_{1}+O(\alpha_{2});\;\;\;\; (\alpha_{2}\ll \alpha_{1})$.}. 
Thus the amplitudes of the Fig.7 type, where the gluon $k'$ is
emitted one time by the transverse gluon and another time by the
longitudinal one, cancel each other and we come back to the ladder
configuration with the ordering eqs.(2.30), (2.31).\\ \\

\section{Infrared Evolution Equations}
\setcounter{equation}{0}

In this section we construct the infrared evolution equations which are
necessary for the calculation of $T_3$ and $g_1$.
We shall follow ~\cite{BER}, and we begin with the amplitude $T_3$ which,
following the discussion of the previous section, consists of the two
components:
\beqn
T_3= \left( \begin{array}{c}
             T_3(\gamma^* g) \\ T_3(\gamma^* q)
               \end{array} \right)
\eeqn
(from now on it will be understood that we consider the singlet part only,
and we suppress the subscript ``S``.) The structure function then follows from
the relation (2.3), and we have to take into account both 
DL-contributions and $i\pi $ -terms. We write $T_3$ as a Mellin transform:
\beqn
T_3=\int_{-i\infty}^{i\infty} \frac{d \omega}{2\pi i}
(\frac{s}{\mu^2})^\omega \xi_(\omega) R(\omega, y),
\label{f28}
\eeqn
where $R(\omega,y)$ is a two-component vector, defined on analogy to (3.1).
The signature factor $\xi(\omega)=$ is:
\be
=\frac{e^{-i\pi \omega}-1}{2}\approx \frac{-i\pi \omega}{2},
\label{f29}
\ee
and 
\be
y=\ln ( \frac{Q^2}{\mu^2}).
\label{f30}
\ee
Thus
\be
- \mu^2 \frac{\partial R}{\partial \mu^2}=
(\omega+\frac{\partial}{\partial y}) R.
\label{f31}
\ee
Eq.( \ref{f31}) represents the left-hand side of the IREE for $T_3$ which 
is illustrated in Fig.8. The right hand side is obtained from the observation 
that the dependence upon the cutoff $\mu$ resides in the intermediate state
with lowest virtuality (Fig.8): the $\mu$-derivative of the amplitudes 
are equal to R times quark or gluon scattering amplitudes with the external 
legs having transverse momenta close to $\mu$. 
We are thus lead to the definition of a quark-quark scattering amplitude
for which we use an integral representation of the form (3.2). 
The partial wave will be denoted by $F_{qq}$. More general,
we introduce the four amplitudes $F_{qq}$, $F_{qg}$,
$F_{gq}$, and $F_{gg}$, and in analogy with (2.28) 
we combine them into the two by two matrix $F_0$ 
\beqn
F_0= \left( \begin{array}{cc}
          F_{gg} & F_{qg} \\
          F_{gq} & F_{qq}
          \end{array} \right)
\eeqn
(it is the analogue to $f_0^{(-)}$ in ~\cite{BER}). In terms of this $F_0$, 
the vector evolution equation for $R$ becomes (Fig.8):
\beqn
(\omega + \frac{\partial}{\partial y}) R = \frac{1}{8 \pi^2}  F_0 R.
\eeqn
In analogy with ~\cite{BER,KL} the evolution equation of $F_0$ has the form (Fig.9a):
\beqn
F_0(\omega)=\frac{g^2}{\omega} M_0 - \frac{g^2}{2 \pi^2 \omega^2}  
             G_0 F_8(\omega) 
               +\frac{1}{8 \pi^2 \omega} F_0(\omega)^2.
\eeqn
Here we have used the the matrices $M_0$ and $G_0$ defined in the 
previous section.
The second term on the rhs of (3.8) corres
ponds to the gluon bremsstrahlung 
diagrams: in analogy to the matrix $F_0$ which carries color zero we 
define the matrix $F_8$ of color octet amplitudes (it is the analogue to $f_8^{(+)}$ in 
~\cite{BER}). These amplitudes satisfy the evolution equation similar to
(3.8):
\beqn
F_8=\frac{g^2}{\omega} M_8 + \frac{g^2 C_A}{8 \pi^2 \omega} 
           \frac{d}{d \omega} F_8(\omega)
          +\frac{1}{8 \pi^2 \omega} F_8(\omega)^2.
\eeqn 
The matrix $M_8$ is taken from the previous section, and the color
factor $C_A$ in front of the second term on the rhs is the analogue of the
matrix $G_0$ in (3.8). The difference between $C_A$ in (3.9) and $G_0$ in
(3.8) is due to the fact that
for the positive signature amplitude $F_8$ the sum of the two
bremsstrahlungs diagrams (illustrated in Fig.9b) is independent of the 
type of the incoming partons,
and the matrix of color factors $G_8$ becomes $C_A$ times the unit matrix.
\\ \\

\section{Solution of the evolution equation}
\setcounter{equation}{0}
The solution for $g_1$ is obtained by solving eq.(3.9) for $F_8$, then
eq.(3.8) for $F_0$ and eq.(3.7) for $R$ (the latter makes use of the Born
approximation $R_B$ as initial condition for $R$), and finally 
inserting $R$ into (3.2). The final answer for $T_3$ is the two component 
vector:
\beqn
T_3 = \int \frac{d \omega}{2 \pi i} 
          \left( \frac{1}{x} \right)^{\omega} \xi(\omega)
          \left( \frac{Q^2}{\mu^2} \right)^{F_0/8\pi^2}
                       \frac{1}{\omega - F_0 /8 \pi^2} R_B
\eeqn
Let us go through these steps in somewhat more detail.\\ \\
We begin with the equation for $F_8$. We first diagonalize the Born term,
i.e. the matrix $M_8$. The eigenvalues are:
\beqn
\lambda_8^{(\pm)} = \frac{2C_A - 1/2N}{2} 
              \pm \frac{1}{2}\sqrt{ (2C_A-1/2N)^2 - 4 C_A T_f }.
\eeqn
Let $e^{(+)}$ and $e^{(-)}$ denote the two eigenvectors of $M_8$  
and $E_{8}=(e^{(+)},e^{(-)})$ the matrix composed of them. Then
\beqn
M_8 = E_8 \hat{M_8} E_8^{-1},
\eeqn
where $\hat{M_8}= diag(\lambda_8^+, \lambda_8^-)$. Consequently, eq.(3.9)
becomes diagonal if we transform to $\hat{F_8}$:
\beqn
F_8 = E_8 \hat{F_8} E_8^{-1}.
\eeqn
The solutions for the two components of $\hat{F_8}$ are:
\beqn
\hat{F_8}^{\pm} = N g^2 \frac{\partial}{\partial \omega}
               \left(  \ln e^{z^2/4} D_{p_{\pm}} (z) \right),
\eeqn
where
$D_p$ denotes the parabolic cylinder function with
\beqn
p_{\pm} = \frac{\lambda_8^{(\pm)}}{N}
\eeqn
and
\beqn
z=\frac{\omega}{\omega_0}, \,\,\,\, \omega_0 = \sqrt{Ng^2/8 \pi^2}.
\eeqn
\\ \\
With this solution for $F_8$ we return to the evolution equation (3.8) for 
$F_0$ which is solved by the (matrix-valued) expression:
\beqn
\frac{1}{4\pi^2 \omega} F_0 = 1 - \sqrt{1 - \frac{g^2}{2(\pi \omega)^2} M_0
               +\frac{g^2}{4\pi (\pi \omega)^3} G_0 F_8 }
\eeqn
This has to be substituted into the final formula (3.2) for $T_3$. The
leading behavior at small x comes from the right-most singularity in the
$\omega$ plane. Similar to the nonsinglet case, this singularity is due
to the vanishing of the square root in (4.8), i.e. we need to determine
the zeroes of the eigenvalues of the matrix under the square root. Similar
to the discussion after (4.2), the diagonalization of this matrix is done
by the matrix $E_{0}$ consisting of the two eigenvectors $e_0^{(+)}$ and
$e_0^{(-)}$ of the matrix under the square root. \\ \\
Let us first return to the eigenvalues of $M_8$ determining the 
asymptotic behavior of $F_8$. For four active flavors the eigenvalues
are:
\beqn
\lambda_8^{\pm} = 2.92 \pm 1.88 = \left( \begin{array}{c}
                               4.8 \\ 1.04 
                                 \end{array}  \right).
\eeqn
Neglecting fermions one obtains $\lambda=6$, whereas the pure fermionic case
corresponds to $\lambda=-1/6$. Obviously, the larger of the two eigenvalues
is not too far from the gluonic case.\\ \\
The accurate values have to be found from a numerical computation of the 
parabolic cylinder function. Before we quote the results of such a 
calculation it may be instructive to discuss a few approximations.
Turning to the square root expression in
(4.8), we are searching values of $\omega$ (or z) for which one of the two
eigenvalues of the matrix under the square root in (4.8)  becomes zero. 
To obtain a first estimate, we simply neglect
the term containing $F_8$ (the exact calculation will show that this
approximation is not too bad): then we can diagonalize $F_0$ by simply
diagonalizing $M_0$. The eigenvalues are
\beqn
\lambda_0^{\pm}= \frac{1}{2} (4C_A + C_F) \pm \frac{1}{2} 
        \sqrt{ (4C_A+C_F)^2 - 16 C_A C_F -16 C_F T_f}
\eeqn
with the corresponding eigenvectors:
\beqn
e^{(+)}=\left( \begin{array}{c} 1 \\ x^{(+)} \end{array} \right),\,\,\, &
x^{(+)}=\frac{\lambda_0^+ - M_{0\,\,11}}{M_{0\,\,12}} = 
            \frac{\lambda_0^+ - 4C_A}{-2T_f} \\
e^{(-)}=\left( \begin{array}{c} x^{(-)} \\ 1 \end{array} \right), \,\,\,&
x^{(-)}=\frac{\lambda_0^- - M_{0\,\,22}}{M_{0\,\,21}} =
            \frac{\lambda_0^- - C_F}{2 C_F}.
\eeqn         
For the diagonalization we need the matrix $E_{0}=(e^{(+)}, e^{(-)})$ and
its inverse
\beqn
E_0^{-1} =\frac{1}{1-x^+ x^-} \left( \begin{array}{cc}
1 & -x^- \\ -x^+ & 1 \end{array} \right).
\eeqn 
The two values of z are (for four active flavors):
\beqn
z_s = \left( \begin{array}{c}
            3.81 \\ 1.81
           \end{array}   \right)
\eeqn
(the pure gluonic and fermionic cases give $z=4$ and $z=4/3$ resp.). 
With the eigenvalues
\beqn
\lambda_0^{\pm}=\left( \begin{array}{c} 10.88 \\ 2.46 \end{array} \right)
\eeqn
the matrix $E_0$ is found to be:
\beqn
E_0= \left( \begin{array}{cc} 1 & 0.42 \\ 0.28 & 1 \end{array} \right).
\eeqn
For the partial wave $R(\omega, y)$ in (3.2) we obtain:
\beqn
R(\omega, y) = 
          E_0 \frac{1}{\omega - \hat{F_0}/8 \pi^2} 
    \left( \frac{Q^2}{\mu^2} \right)^{\hat{F_0}/8 \pi^2} E_0^{-1}
\left( \begin{array}{c} 0 \\ 2 e_q^2 \end{array} \right).
\eeqn
Retaining in $\hat{F_0}$ only the leading upper component, we find for the
behavior of $R$ near the square root branch point at $\omega = \omega_s$:
\beqn
R(\omega_s, y) \sim  
\frac{1}{1-0.12} \left( \begin{array}{cc} 1 & -0.42 \\ 0.28 & -0.12 \end{array}
       \right) \frac{2}{\omega_s}
           \left( \frac{Q^2}{\mu^2} \right)^{\omega_s/2}
\left( \begin{array}{c} 0 \\ 2 e_q^2 \end{array} \right).
\eeqn
The fact that the matrix elements in the rightmost column are both negative 
has the important consequence that the leading contribution at small x 
changes the sign relative to the input distribution. 
Finally,
\beqn
g_1^S  = \frac{\omega_s^{3/2}}{8 \sqrt{2 \pi}}
 \frac{\frac{2}{\omega_s}+\ln Q^2/\mu^2}{(\ln (1/x))^{3/2}} 
(\Delta g, \Delta \Sigma) R(\omega_s,Q^2) 
(\frac{1}{x})^{\omega_s} 
\left( 1+O(\frac{\ln^2 Q^2/\mu^2}{\ln 1/x}) \right) 
\eeqn
where the vector $R$ is from (4.18), and the (transposed) vector 
$(\Delta g$, $\Delta \Sigma)$ represents the initial
conditions of the gluon and quark polarized distributions. It 
should be kept in mind that here we have
retained only the leading singularity $\omega_s$ of (4.14). With $\omega_0$
from (4.7) and $\alpha_s=0.18$ we find $\omega_s=z_s \sqrt{\alpha_s N_c/2 \pi}
=1.12$.
\\ \\
Finally, taking into account also $F_8$ we quote the result of a 
numerical calculation. For the larger of the two z-values we find ($n_f=4$):
\beqn
z_s=3.45
\eeqn
(the pure gluonic case would have given $z=3.66$). With $\omega_0$ from (4.7)
and $\alpha_s=0.18$ we find
\beqn
\omega_s = z_s \sqrt{\alpha_s N_c /2 \pi} =1.01,
\eeqn
and with $x^+ = 0.29$, $x^- = 0.43$
\beqn
R(\omega_s, y) \sim  
 \left( \begin{array}{cc} 1.14 & -0.50 \\ 0.33 & -0.14 \end{array}
       \right) \frac{2}{\omega_s}
           \left( \frac{Q^2}{\mu^2} \right)^{\omega_s/2}
\left( \begin{array}{c} 0 \\ 2 e_q^2 \end{array} \right),
\eeqn
i.e. the negative sign persists. The result for $g_1$ becomes:
\beqn
g_1(x,Q^2) = \frac{\omega_s^{3/2}}{8 \sqrt{2 \pi}}
\frac{\frac{2}{\omega_s}+\ln Q^2/\mu^2}{(\ln(1/x))^{3/2}}   
(\Delta g, \Delta \Sigma) R(\omega_s,y) 
(\frac{1}{x})^{\omega_s} \left( 1 + O(\frac{\ln^2 Q^2/\mu^2}
{\ln 1/x}) \right)
\eeqn
with $R$ from (4.22). 
\\ \\
Let us finally compare our result with the fixed order calculations.
The region where we expect our result to coincide with the fixed order
calculation is $\sqrt{\alpha_s} \ll \omega \ll 1$. We therefore
expand our anomalous dimension matrix in powers of $g^2/\omega^2$. To leading
order we obtain:
\beqn
\gamma_{S}^{(0)}= 2 \frac{\alpha_s}{4 \pi}\frac{1}{\omega}
           \left( \begin{array}{cc}
                  4C_A & -2T_f \\
                  2 C_F& C_F
                   \end{array} \right).
\eeqn
It agrees (as we have already mentioned before)
with the singular part of ~\cite{AP,LO}.
The next-to-leading-order matrix has the form:
\beqn
\gamma_{S}^{(1)}= (\frac{\alpha_s}{4 \pi})^2 \frac{1}{\omega^3}
            \left( \begin{array}{cc}
                  32 C_A^2 -16 C_F T_f & -16 C_A T_f -8 C_F T_f \\
                  16 C_A C_F +8 C_F^2 & 4 C_F^2 -16 C_F T_f +\frac{8C_F}{N}
                   \end{array} \right).
\eeqn
After transformation to $x_B$, it agrees with the leading powers on $\ln x$ 
given in eq.(3.65)-(3.68) of ~\cite{MvN}.

\section{Discussion}
\setcounter{equation}{0}

The main result of this paper is the power-like behavior of the flavor singlet 
part of $g_1$ at small x (eq.(4.23)). The (leading) power is by a factor 2.6 
larger than in the 
nonsinglet case. This effect is mainly due to the t-channel gluons states,
which have a much larger color charge (cf.eq.(2.28), (2.29)) than the quarks 
(comparing the pure gluon and the pure fermionic case and neglecting
the nonladder gluons one finds $\omega^{singlet}/\omega^{nonsinglet}
=\sqrt{4C_A/C_F}=3$). As it can be seen from a comparison of (4.14) and (4.20),
the influence of the bremsstrahlung gluon is of the order of $10 \% $.
Comparing the flavor singlet with the nonsinglet, one notices that
the nonladder bremsstrahlungs gluons act in quite different ways: in the 
present 
case, due to the positive sign of the largest matrixelemnt of $M_8$ 
($M_{8 \,\, 11}$), the leading $\omega$-plane singularity 
moves to the left, i.e. the exponent of $1/x$ decreases from $\omega_s=1.12$
to $\omega_s=1.01$. For the flavor nonsinglet, on the other hand, the 
nonladder gluons lead to an increase of the exponent. \\ \\
As to the polarization of the t-channel ladder gluons, 
the following
pattern has emerged. In the unpolarized case, both t-channel gluons are
longitudinally polarized (i.e. they are in the nonsense helicity state), and
the resulting behaviour is $g(x,Q^2) \sim (1/x)^{1+O(\alpha_s)}$. 
In the polarized case, this leading contribution of gluon polarizations
cancels, and the next-to-leading configuration appears: one gluon is still
longitudinally polarized, but the other one has transverse polarization.
The small-x behaviour is now of the form 
$g_1(x.Q^2) \sim (1/x)^{O(\sqrt{\alpha_s})}$. If both gluons are 
transversely polarized, the small-x behaviour is further suppressed: 
$\sim (1/x)^{-1+O(\sqrt{\alpha_s})}$. This contribution has been studied in
~\cite{FK}.\\ \\
Compared to the small-x prediction of the standard GLAP evolution equation,
the main reason for the strong enhancement of our calculation lies in the
different regions of phase space. Whereas in the GLAP case we have strong 
ordering in transverse momenta of the ladder partons, our double 
logarithmic approximation has a much
larger region of integration. A good way of illustrating the difference is the
diffusion in $\ln k_T^2$ which first has been observed for the BFKL
~\cite{BFKL}
Pomeron. When going, within ladder graphs, from one rung ``i`` to the next
rung ``i+1``, ordering in transverse momenta means $k_{Ti} < k_{Ti+1}$. 
In the BFKL approximation, the distribution in $\ln k_{Ti+1}$ is a random walk
distribution, i.e. when climbing up the ladder up to rung ``i+1``, we either
add or subtract, at each step, a fixed steplength $\Delta$. In the present
double logarithmic approximation, on the other hand, the steplength is
not fixed but can vary between zero and $\ln 1/x$. From this qualitative 
argument it is clear that, in the {\it polarized} case, the difference 
between the GLAP small-x behaviour and the double logarithmic prediction
should be larger than, in the {\it unpolarized} case, the difference between
GLAP and the BFKL behaviour.\\ \\
Finally, at least a few words should be said about the phenomenological
applications of our calculation ~\cite{EMR,GRSV,BV}. Of particular interest 
is, of course, the
question at what values of $x$ and $Q^2$ the GLAP extrapolation should be
replaced by the result of this paper. Leaving a more detailed discussion for
future publications, we only want to make a few general remarks. In this paper
we have derived expressions for Greens functions - $G_{q,g}(x,x';Q^2, \mu^2)$ -
which describe the evolution from the initial point $(x',\mu^2)$ to the
final point $(x,Q^2)$. If the initial polarized quark and gluon distribution
do not exhibit any particularly strong variation in $x$, we can start from the
approximation:
\beqn
g_1(x,Q^2) = \Delta \Sigma(\bar{x}, \mu^2) G_{q}(x,\bar{x}; Q^2, \mu^2)
          + \Delta g(\bar{x},\mu^2) G_{g}(x,\bar{x};Q^2,\mu^2) 
\eeqn
where $\bar{x}$ denotes the mean x-value of the initial distributions.
As a general feature, our Greens functions rather strongly depends upon the
momentum scale $\mu^2$.\\ \\
Our analytic expression (4.23) represents only 
the leading term of the full amplitude (4.1). We have checked 
numerically that, 
as a function of x, the full amplitude (4.1) quickly shows the power behaviour
predicted by the leading approximation (due to the factor $\ln 1/x$ 
there is, however, some difference in the overall normalization). 
Furthermore, as (4.22) suggests, at small x
the gluon will dominate over the quark component. A numerical calculation
shows that this, again, happens rather rapidly: the ratio of gluon and quark 
Greens functions $G_g/G_q$ approaches its asymptotic value 3.45
approximately at $x=10^{-3}$.\\ \\
An important aspect of our result (4.23) is the sign structure.
Our formula predicts that, at small x, the sign of $g_1$ should be opposite
to that of the initial distributions.
The experimental data \cite{data}  indicate 
that at small x the proton function $g_1^p(x)$ is positive, and even the 
neutron  function $g_1^n$ for $x<0.01$ shows the tendency to change the sign 
and become positive. So, at first sight, there seems to be a problem.
A possible way of resolving this conflict could be the following.
Whereas the value of $\Delta\Sigma$ is known (and positive), we do not know
the initial value of $\Delta g$. In ref.\cite{J} it was shown that in a 
resonable model of the nucleon (non-relativistc quark or bag-model) 
this value may be negative, of about $\Delta g\sim -0.4$. Taking into 
account that at small x the gluon Greens function will dominate over 
the quark component, the structure function $g_1$ may very well become
positive - both for the proton and the neutron - and thus be consistent
with the behavior observed in the data.\\ \\
All these arguments are clearly very qualitative, and a more 
detailed quantitative analysis is of great importance.
For the nonsinglet case, a first
attempt has been made in ~\cite{EMR} to estimate numerically how large
the deviations of our double log formula from the usual GLAP predictions
might be. Clearly any such estimate will strongly depend upon the input
distribution. In ~\cite{EMR} it was concluded that the deviations may be very 
large (up to an order of magnitude) but the difference between the nonsinglet
$g_1(x,Q^2)$ and $f_1(x,Q^2)$ will not
exceed 15 \% within the HERA kinematical range. A somewhat different 
analysis has been performed in ~\cite{BV}.
The situation with the singlet contribution will be more
involved: not only is the power generated by the double logs larger than
in the nonsinglet case, also the sign structure of the leading term 
(as mentioned 
in the previous section) will make the result more dependent upon the input.
Clearly a careful study in this direction is strongly needed. On the longer
run, also an improvement in accuracy of our double logarithmic
approximation should be persued.\\ \\
{\bf Acknowledgements:} We are grateful to L.N.Lipatov for useful 
discussions. Two of us (B.E. and M.R.) wish to 
thank DESY for the kind hospitality, and the Volkswagen Stiftung for 
financial support.\\ \\
{\bf Figure Captions:}\\ \\
Fig.1: Born approximation for $M^S$.\\ \\
Fig.2: First corrections to $M^S$: gluon rungs in the fermion ladder.\\ \\
Fig.3: Lowest order contribution with t-channel gluons.\\ \\
Fig.4: The first gluon rung.\\ \\
Fig.5: Calculation of the gluon rung in Fig.4.\\ \\
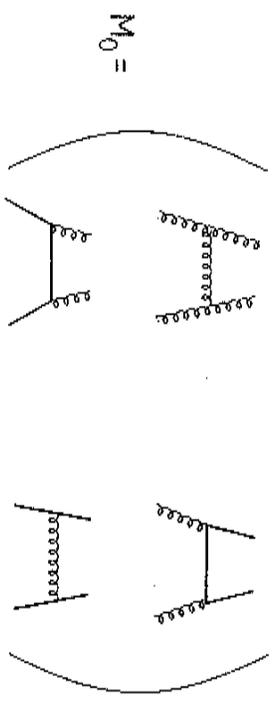
Fig.6: Matrix notation in Eq.(2.17).\\ \\
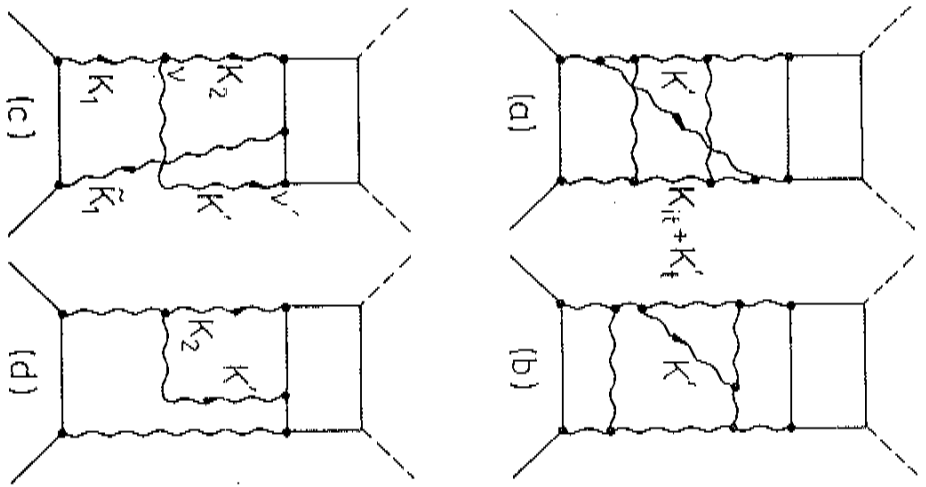
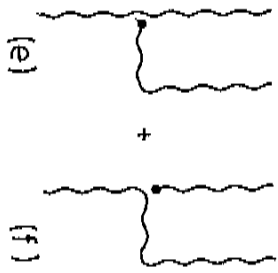
Fig.7: Nonladder graphs.\\ \\
Fig.8: Illustration of the infrared evolution equation (3.7).\\ \\
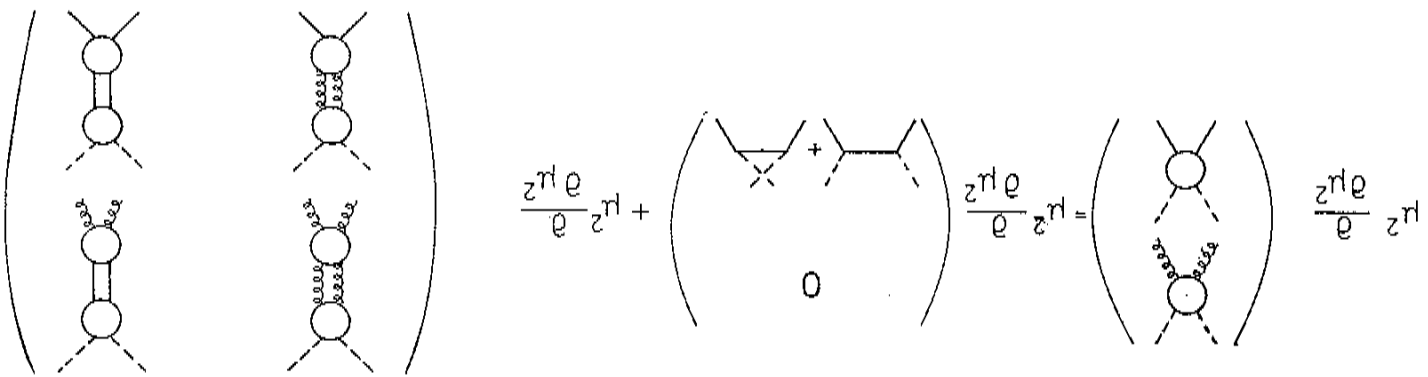
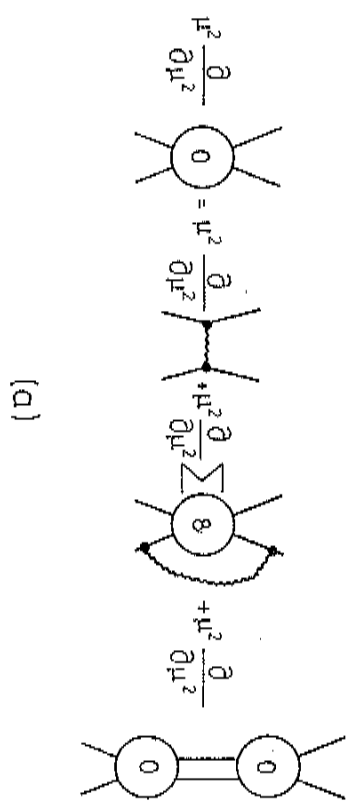
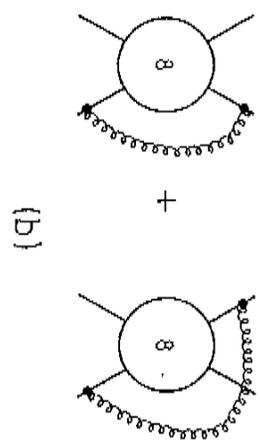
Fig.9: (a) Illustration of the evolution equation (3.9); 
(b) origin of the color factor in front of the second term in (3.9).  

\end{document}